\documentstyle[12pt]{article}

\advance\textheight by \topskip
\begin{document}
\tolerance=100000
\thispagestyle{empty}
\setcounter{page}{0}

\newcommand{\mathrm}{\rm}
\newcommand{\be}{\begin{equation}}
\newcommand{\ee}{\end{equation}}
\newcommand{\br}{\begin{eqnarray}}
\newcommand{\er}{\end{eqnarray}}
\newcommand{\ba}{\begin{array}}
\newcommand{\ea}{\end{array}}
\newcommand{\bi}{\begin{itemize}}
\newcommand{\ei}{\end{itemize}}
\newcommand{\bn}{\begin{enumerate}}
\newcommand{\en}{\end{enumerate}}
\newcommand{\bc}{\begin{center}}
\newcommand{\ec}{\end{center}}
\newcommand{\ul}{\underline}
\newcommand{\ol}{\overline}
\newcommand{\eeffwz}{$e^+e^-\rightarrow f\bar f' W^\pm Z^0$}
\newcommand{\eeffzz}{$e^+e^-\rightarrow f\bar f Z^0Z^0$}
\newcommand{\eeffz}{$e^+e^-\rightarrow f\bar f Z^0$}
\newcommand{\eezz}{$e^+e^-\rightarrow Z^0Z^0$}
\newcommand{\ffzz}{$f\bar f Z^0 Z^0$}
\newcommand{\ffwz}{$f\bar f' W^\pm Z^0$}
\newcommand{\ffz}{$f\bar f Z^0$}
\newcommand{\eehz}{$e^+e^-\rightarrow H^0Z^0$}
\newcommand{\uub}{$ u\bar u$}
\newcommand{\ddb}{$ d\bar d$}
\newcommand{\ssb}{$ s\bar s$}
\newcommand{\ccb}{$ c\bar c$}
\newcommand{\bbb}{$ b\bar b$}
\newcommand{\ttb}{$ t\bar t$}
\newcommand{\eeb}{$ e^+ e^-$}
\newcommand{\mumub}{$ \mu^+\mu^-$}
\newcommand{\tautaub}{$ \tau^+\tau^-$}
\newcommand{\veveb}{$ \nu_e\bar\nu_e$}
\newcommand{\vmvmb}{$ \nu_\mu\bar\nu_\mu $}
\newcommand{\vtvtb}{$ \nu_\tauu\bar\nu_\tau $}
\newcommand{\lra}{\leftrightarrow}
\newcommand{\ar}{\rightarrow}
\newcommand{\sm}{${\cal {SM}}$}
\newcommand{\MH}{$M_{H^0}$}
\newcommand{\Dir}{\kern -6.4pt\Big{/}}
\newcommand{\Dirin}{\kern -10.4pt\Big{/}\kern 4.4pt}
\newcommand{\DDir}{\kern -7.6pt\Big{/}}
\newcommand{\DGir}{\kern -6.0pt\Big{/}}
\def\Ord{\buildrel{\scriptscriptstyle <}\over{\scriptscriptstyle\sim}}
\def\OOrd{\buildrel{\scriptscriptstyle >}\over{\scriptscriptstyle\sim}}
\def\pl #1 #2 #3 {{\it Phys.~Lett.} {\bf#1} (#2) #3}
\def\np #1 #2 #3 {{\it Nucl.~Phys.} {\bf#1} (#2) #3}
\def\zp #1 #2 #3 {{\it Z.~Phys.} {\bf#1} (#2) #3}
\def\pr #1 #2 #3 {{\it Phys.~Rev.} {\bf#1} (#2) #3}
\def\prep #1 #2 #3 {{\it Phys.~Rep.} {\bf#1} (#2) #3}
\def\prl #1 #2 #3 {{\it Phys.~Rev.~Lett.} {\bf#1} (#2) #3}
\def\mpl #1 #2 #3 {{\it Mod.~Phys.~Lett.} {\bf#1} (#2) #3}
\def\rmp #1 #2 #3 {{\it Rev. Mod. Phys.} {\bf#1} (#2) #3}
\def\xx #1 #2 #3 {{\bf#1}, (#2) #3}
\def\preprint{{\it preprint}}

\begin{flushright}
{\large DFTT 40/95}\\
{\large DTP/95/58}\\
{\rm June 1995\hspace*{.5 truecm}}\\
\end{flushright}

\vspace*{\fill}

\begin{center}
{\Large \bf The intermediate-mass \sm\ Higgs boson at the NLC:
reducible and irreducible backgrounds to the Bjorken production
channel\footnote{Work supported in part by Ministero
dell' Universit\`a e della Ricerca Scientifica.\\[4. mm]
E-mails: Moretti@to.infn.it;
Stefano.Moretti@durham.ac.uk.}}\\[2.cm]
{\large Stefano Moretti\footnote{Address
after September 1995: Cavendish Laboratory,
University of Cambridge,
Madingley Road,
Cambridge, CB3 0HE, U.K.}}\\[0.5 cm]
{\it Dipartimento di Fisica Teorica, Universit\`a di Torino,}\\
{\it and I.N.F.N., Sezione di Torino,}\\
{\it Via Pietro Giuria 1, 10125 Torino, Italy.}\\[0.5cm]
{\it Department of Physics, University of Durham,}\\
{\it South Road, Durham DH1 3LE, United Kingdom.}\\[0.75cm]
\end{center}

\vspace*{\fill}

\begin{abstract}
{\normalsize
\noindent
Both the reducible and irreducible backgrounds to the Higgs production
channel \eehz\ at a Next Linear Collider (NLC)
are studied, for the Standard Model (\sm) Higgs boson in the
intermediate-mass range. A phenomenological analysis that does not
exploit any form of tagging on the Higgs decay products
is assumed.}
\end{abstract}

\vspace*{\fill}
\newpage
\subsection*{1. Introduction}

The Higgs mechanism of spontaneous symmetry breaking of the
electroweak interactions is a cornerstone of the Standard Model (\sm).
It can explain why in nature some fundamental
particles (i.e. leptons, quarks, the $Z^0$ and $W^\pm$ gauge bosons) have
a non-zero mass. A consequence of the mechanism is
that it predicts the existence of a ${\cal
{CP}}$-even neutral scalar boson (i.e. the Higgs boson $H^0$), which
has not been observed yet.
Therefore the discovery of such a particle is crucial in order to
assess the correctness of the whole model.\par
Experimental (for the lower bound, see \cite{limSM}) and theoretical
(for the upper
bound, see \cite{unitarity}) analyses have established that the Higgs boson
mass should be in the range 64 GeV $\Ord M_{H^0}\Ord $ 700 GeV.
Depending on the value of \MH,
many studies on the feasibility of its detection and on the
possibilities of measuring its parameters (i.e. other than the mass
$M_{H^0}$:
the width $\Gamma_{H^0}$, the spin and parity, the couplings to the other
particles, etc ...) have been carried out, both for hadron
\cite{hadron} and
$e^+e^-$ colliders \cite{epem}.\par
While the mass
intervals \MH\ $\Ord 100$ GeV and \MH\ $\OOrd 2M_{W^\pm}$
should be easily covered by LEP II and LHC,
respectively, the remaining
range (intermediate-mass), which is beyond the possibilities of LEP
II, appears much more difficult since for values of \MH\ in this interval the
Higgs boson mainly decays to \bbb-pairs, a signature which has
 a huge QCD background  at the
LHC.
Nevertheless, some important results concerning the possible
detection of an intermediate-mass Higgs at the CERN $pp$ collider have
been achieved.
First of all, the fact that the $H^0$ can decay also to $Z^{0}Z^{0*}$-pairs,
in which one of the two gauge bosons is largely
off-shell, allows for the Higgs detection in the `gold
plated' channel $H^0\ar 4\ell$ already starting from $M_{H^0}\approx 130$ GeV.
Secondly, if $M_{Z^0}\Ord M_{H^0}\Ord 130$ GeV, it is possible in principle
to exploit two different strategies. Either one can search for rare
non-hadronic Higgs decays (i.e. $H^0\ar \gamma\gamma$),
or instead detect the main decay
channel (i.e. $H^0\ar b\bar b$) by resorting to techniques of
$b$-flavour identification ($b$-tagging).\par
However, it should be remembered that these two latter approaches
rely on the fact
that a large luminosity by the CERN $pp$ machine and/or that
very high tagging efficiencies
(in photon resolution and in $b$-tagging respectively)
can be achieved by  the LHC detectors. In fact, a very detailed
study \cite{Froidevaux} has recently claimed that even for optimistic
$b$-tagging performances
and integrated luminosities of the order
$\int{\cal L}dt=10^4$ pb$^{-1}$, the $H^0\ar b\bar b$
signal cannot be cleanly extracted from the background. Nevertheless,
after a few years of running at the LHC with  a Center-of-Mass (CM)
energy of 14 TeV, this channel might be the
best way to probe the region 80 GeV $\Ord M_{H^0}\Ord$ 100 GeV,
whereas (if a higher
luminosity can be achieved) the $H^0\ar\gamma\gamma$ signature is
better for 100 GeV $\Ord M_{H^0}\Ord$ 130 GeV.
\par
Now, if we consider that other than in the `detection' of the \sm\ Higgs
boson we are interested in measuring its parameters in detail
(because of, e.g. the implications that some of these could have for
the existence of
possible Supersymmetric extensions of the \sm),
the importance of a Next Linear Collider (NLC) is immediately apparent.
The advantage of such a machine
(where two electron-positron beams linearly collide at a CM energy
$\OOrd$ 300--350 GeV) with respect to a hadron collider is that here
the QCD background is drastically reduced, and one can easily exploit
in the intermediate-mass range the main decay channel $H^0\ar b\bar
b$.\par
For a first stage NLC (with $\sqrt s\approx$ 300--350 GeV) the main
production mechanism of an intermediate-mass Higgs is the Bjorken
reaction $e^+e^-\ar Z^{0*} \ar Z^0H^0$ \cite{Bjorken}, which dominates
over the $W^\pm W^\mp$ and $Z^0Z^0$ fusion channels
$e^+e^-\rightarrow \bar\nu_e\nu_eW^{\pm*}
W^{\mp*}(e^+e^-Z^{0*}Z^{0*})\rightarrow\bar\nu_e\nu_e
(e^+e^-)H^0$ \cite{fusion}.
At larger CM energies ($\sqrt s\OOrd 500$ GeV) it is the other way
round.\par
Because of the crucial role that a NLC could have
for detecting and studying a Higgs boson with an intermediate-mass,
it is then extremely important to exploit all possible search strategies
and to carefully know all the corresponding
backgrounds (both
reducible and irreducible), which could, in principle, prevent
measuring the parameters of the $H^0$ with the needed accuracy.\par
It is the purpose of this paper to study the characteristics of the
signal and of all possible backgrounds for
a `Bjorken Higgs' in the intermediate-mass range  produced at the NLC.
We will one assume as a
search strategy the method of calculating the mass recoiling
against the $Z^0$ (missing-mass analysis) \cite{GHS}, without
selecting any of the specific Higgs decay channels but instead
considering them altogether in a sort of `inclusive' analysis.
In this kind of approach the $Z^0$ is
most conveniently reconstructed by its \eeb\ and \mumub\ decay modes,
but also hadronic $Z^0$ decays, even the case $Z^0\ar b\bar b$ (with
$b$--tagging), can be used. For our convenience,
we will take in the numerical computations the $Z^0$ to be on-shell.\par
Such a strategy has the useful
feature of being completely independent of assumptions
about the $H^0$ decay modes but requires only tagging only the decay
products
of the $Z^0$ produced in the two-to-two body Bjorken reaction.
Therefore, it demands less experimental effort, with respect to the case in
which `exclusive' Higgs channels are considered, whether one attempts
the full kinematic reconstruction of the reaction $e^+e^-\ar Z^0H^0$
(via the decays $Z^0H^0\ar$ jets and/or leptons), or one directly
reconstructs the invariant mass from the the jets (through the decays $H^0\ar
b\bar b$ and/or $W^{\pm *}W^\mp$) \cite{GHS}.\par
On the contrary, in this `inclusive' approach,
it is necessary then not only to compute the rates for
all possible decay channels of the Higgs boson and the corresponding
irreducible backgrounds, but also the ones of reactions
producing a $Z^0$ in association
with additional particles faking possible Higgs decays,
which appear in the invariant mass
recoiling against the primary $Z^0$ but
do not contribute to the signal spectrum.
Therefore,
reducible backgrounds such as, e.g. the
processes $e^+e^-\ar Z^0 q\bar q$, for light flavours $q=u,d,s,c$
should now be considered. This generally acts in the direction of
reducing the significance of the signal, as the Higgs boson
practically never decays to $q\bar q$-light pairs, whereas the
contribution from events $e^+e^-\ar Z^0Z^{0*}+Z^0\gamma^*\ar Z^0q\bar
q$ is expected to be quite large \cite{GHS}.\par
In general, however, the use of missing-mass techniques
applied to the process $e^+e^-\ar Z^0H^0$ is quite powerful,
as it provides a very efficient experimental technique to detect the Higgs
scalar or to rule out this particle with certainty.
Moreover, it
is particularly important, e.g. in the Minimal Supersymmetric
extension of the Standard Model (${\cal{MSSM}}$), if $M_\chi<M_{h}/2$
(where
$\chi$ represents a neutralino and $h$ the lightest ${\cal{MSSM}}$
 neutral scalar
boson). In fact, in this case, the `invisible' decay
$h\ar\chi\chi$ is the dominant decay channel in the intermediate range
of $M_h$ and cannot be directly measured. Nevertheless,
by missing-mass analyses, the signal $e^+e^-\ar
Zh\ar (\ell^+\ell^-)(\chi\chi)$ ($\ell=e$ or $\mu$) should clearly
appear as a peak in the recoiling mass distribution \cite{Janot}.\par
The plan of this paper is as follows. In Section 2 we give brief
details of the computations, as well as the
input numerical parameters. In Section 3 we discuss the results,
whereas in Section 4 we present the conclusions.

\subsection*{2. Calculation}

We are interested in an analysis that does not perform any
tagging on the decay products of a \sm\ Higgs boson produced via the
Bjorken bremsstrahlung
reaction \eehz. We shall focus on the intermediate-mass range
$M_{Z^0}\Ord M_{H^0}\Ord 2 M_{W^\pm}$, where the quantitatively
significant Higgs decay channels are into \bbb-,
$W^{\pm *}W^\mp$- and $Z^{0*}Z^0$-pairs,
giving the signatures $jj$ (no flavour identification of $b$--quarks
is assumed in our analysis),
$jjW^\mp$, $\ell\nu_\ell W^\mp$,
$jjZ^0$, $\ell\bar\ell Z^0$ and
$\nu_\ell\bar\nu_\ell Z^0$, with all possible subsequent
decays of the on-shell gauge bosons. Thus we are forced to
study the complete processes
\be\label{proc1}
e^+e^-\rightarrow f\bar f Z^0,
\ee
\be\label{proc2}
e^+e^-\rightarrow f\bar f' W^\pm Z^0,
\ee
\be\label{proc3}
e^+e^-\rightarrow f\bar f Z^0Z^0,
\ee
($f=\ell,\nu_\ell$ and $q$, with $\ell=e,\mu,\tau$ and $q=u,d,s,c,b$). These
include at tree-level all the relevant (both reducible and
irreducible) backgrounds to the Bjorken reaction,
followed by the Higgs decay into the above
channels. For simplicity,
and since the final results would not significantly change, we
neglect here the case of the $H^0$ decaying into $gg$-pairs
through loops of heavy quarks and of the
non-resonant diagrams entering into the
processes $e^+e^-\ar Z^0+n~{\rm{jets}}$ (with $n\ge2$), which are at least
${\cal O}(\alpha^3_W \alpha_s^2)$ suppressed (i.e. a factor of
$\alpha_W \alpha_s^2$ if compared to the $Z^0H^0$ signal, especially
if all the jets are very well separated).
In addition,  Higgs decays
into $\gamma\gamma$ and $Z^0\gamma$ can be safely neglected,
since they contribute only at the level  ${\cal O}(10^{-3})$. At the same
time, it is not necessary to compute their backgrounds in
$e^+e^-\ar Z^0\gamma\gamma$ and $e^+e^-\ar Z^0Z^0\gamma$ events,
if one asks that the mass recoiling against the $Z^0$ does not contain
very hard photons.
\par
Concerning possible backgrounds arising from the $e^+e^-\ar t\bar t\ar
b\bar b W^+W^-$ top-pairs production and decay,
these should be drastically suppressed if we assume for the top mass
a value $m_t\OOrd 175$ GeV \cite{top} and to tag $Z^0\ar\ell^+\ell^-$
(with $\ell=e$ or $\mu$). In fact, at $\sqrt s=300$
GeV the $t\bar t$-threshold is far away, whereas at $\sqrt s=500$
GeV, if one takes, e.g. $m_t=180$ GeV, then the total rate is only
$\approx\sigma(e^+e^-\ar t\bar t)\times
[BR(W\ar\ell\nu_\ell)]^2\approx 7$ fb \cite{eezh1}.
If $Z^0\ar jj$ or $b\bar b$ (with $b$--tagging),
the $t\bar t$ background
would deserve a more detailed treatment that we are not performing here.
However, we expect even this case to be manageable, e.g.
by exploiting the fact that the $Z^0$ produced via the two-body Bjorken
reaction is `practically'
mono-energetic (with $E_{Z^0}\approx E_{\rm{ave}}=
(s-M_{H^0}^2+M_{Z^0}^2)/2\sqrt s$). In fact, this is true apart from
photon bremsstrahlungs off $e^+e^-$-lines (i.e. Initial State
Radiation, ISR). Nevertheless, since the mean $e^+e^-$ CM
energy loss $\delta_{\sqrt s}$ due to ISR is, e.g.
$\approx5\%$ at $\sqrt s=500$ GeV \cite{ISR}, one can
choose a window wide
enough ($\approx \delta_{\sqrt s}\times\sqrt s$) to prevent
complications
due to such effects\footnote{The inclusion
of Linac energy spread and beamsstrahlung should not drastically
change this strategy, at least for the `narrow' D--D and TESLA
collider designs (see ref.~\cite{ISR}).}.
Therefore we expect the cut, say, $|E_{Z^0}-E_{\rm{ave}}|<12.5$ GeV
to be quite efficient in reducing the numbers of $t\bar t$ events
around the $H^0$-peak, as it has been demonstrated in a similar context
in ref.~\cite{eezh2}. In addition, and contemporaneously,
one can always require that $M_{jj(b\bar b)}\approx M_{Z^0}$, in order to
enforce the above $E_{Z^0}$ cut,  improving the mass resolution
as much as needed, depending on the size of the energy losses by
ISR\footnote{We
would also like
to stress here how this procedure should make it unnecessary
to use the veto $M_{jj(b\bar b)j}\ne m_t$ suggested in
ref.~\cite{Orange3}, which would imply tagging three particles,
thus spoiling the attractiveness of this
analysis (which only requires tagging the two decay products of the
`Bjorken $Z^0$').}.
For these reasons
we do not study here, among the background
processes, the $t\bar t$-production and decay.\par
Both the QED radiative
corrections and the genuine weak ones
to the Bjorken process have been computed \cite{bjcorr}. However,
since the backgrounds evaluated here are at tree-level, for consistency,
we use the lowest order rates. In addition, these corrections
 are known to be well
under control.
\par
To give an idea of the complexity of the computations, we show
in fig.~1a-c
all the Feynman diagrams describing at tree-level
processes (\ref{proc1})--(\ref{proc3}) respectively, for $f^{(')}\ne\nu_e,e$.
The cases $f^{(')}=\nu_e,e$ are even more complicated, since they
include also diagrams in which
the incoming electron/positron lines are directly connected to the
final states, and are not shown here.
The matrix elements for the three above reactions
have been computed using the method of ref. \cite{hz},
the {\tt FORTRAN} codes we have written and optimised have been checked for BRS
invariance \cite{BRS} and compared to the corresponding MadGraph/HELAS
outputs \cite{tim}.\par
In order to keep the interplay between the various
resonances,
which appear in the integration domains of the final states
in (\ref{proc1})--(\ref{proc3})  when all tree-level
contributions are kept into account, under control,
we have adopted the technique \cite{eezh1,eezh2} of splitting the corresponding
Feynman amplitudes squared into
a sum of different (non-gauge-invariant) terms and then integrating
each according to its resonant structure.
We will not discuss this in any detail, instead we
refer the reader
to the cited papers. Here, we only want to stress that the `Bjorken
diagrams' giving the signals are the numbers 5 in fig.~1a, 29 in fig.~1b and
18 in fig.~1c\footnote{Even though also the graphs number 27, 28 (11, 12)
in fig.~1b(c), and 17 in fig.~1c
are namely Higgs `Bjorken diagrams', in this case the $H^0$ goes
either into $f\bar f$-pairs, followed by $W^\pm/Z^0$-bremsstrahlung, or
into on-shell $Z^0Z^0$-pairs: decays that are strongly suppressed
or that take place above the
range we are interested in here respectively. However, the first kind
of
graphs (i.e. with $W^\pm/Z^0$-bremsstrahlung)
are properly included in the $H^0\ar f\bar f' W^\pm(f\bar f
Z^0)$ resonance (see ref.~\cite{eezh2} for more details).}, and that
when in the next section we speak of the `missing-mass
distribution' we
mean the sum of the differential cross sections corresponding to
the three processes, each of which is obtained by
summing together the non-gauge-invariant `cross sections'. In that way,
the invariance is perfectly recovered in the end \cite{eezh1,eezh2}.
Obviously, the three above processes do not interfere at all,
and they are computed separately.\par
The multi-dimensional integrations over the phase spaces have been
performed numerically using VEGAS \cite{Vegas}.
The following values of the parameters have been adopted:
$M_{Z^0}=91.1$ GeV, $\Gamma_{Z^0}=2.5$ GeV,
$M_{W^\pm}\equiv M_{Z^0}\cos(\theta_W)\approx80$ GeV,
$\Gamma_{W^\pm}=2.2$ GeV, and $\sin^2 (\theta_W)=0.23$.
For the fermions:
$m_\mu=0.105$ GeV, $m_\tau=1.78$ GeV, $m_s=0.3$ GeV, $m_c=1.4$ GeV, $m_b=4.25$
GeV. All neutrinos and the first generation of quarks/leptons has
been
considered massless: i.e.
$m_{\nu_e}=m_{\nu_\mu}=m_{\nu_\tau}=m_e=m_u=m_d=0$.
The electromagnetic coupling
constant has been set equal to 1/128. For
the Higgs width (i.e. $\Gamma_{H^0}$) we have adopted the tree-level
expression corrected for the running of the quark masses in the vertices
$H^0q\bar q$ (for $q=s,c$ and $b$). They have been
 evaluated at the scale $\mu=M_{H^0}$
\cite{running}. Therefore, in order to be consistent we have used the
same running masses in the corresponding vertex of the production
processes here considered.
Finally, we have avoided
adopting any form of Narrow Width Approximation (NWA), i.e.
the procedure of separately computing the on-shell production
times the branching fractions,
into the final state of processes (\ref{proc1})--(\ref{proc3}),
of the various intermediate particles appearing in the diagrams of figs.~1-3.
Only the final
state $Z^0$'s and $W^\pm$'s are considered on-shell, as for them
we have included neither the effects of their finite width
nor those of their decays. Also, the ISR
\cite{ISR} was not included.
However, we are confident that properly keeping into account
all these aspects would not affect our conclusions.\par

\subsection*{3. Results}

A careful study concerning intermediate-mass Higgs searches at 300--500
\eeb\ linear colliders, for a $H^0$
produced via the Bjorken process, including also a missing-mass analysis
when no assumption on the Higgs decay modes is done,  was presented in
ref.~\cite{GHS}. In that paper, only the \eezz\
background was considered. More recently, a few works studying
`exclusive' signals (i.e. when the decay channels of
the Higgs boson are separately considered), and corresponding backgrounds,
have been completed. Ref.~\cite{BCDKZ} studied the channels
$H^0Z^0\ar (b\bar b) (\mu^+\mu^-), (W^\pm W^{\mp *})(\mu^+\mu^-),
(Z^0Z^{0*}) (\mu^+\mu^-)$, and the backgrounds $e^+e^-\ar Z^0Z^{0*},
Z^0\gamma^*,\gamma^*\gamma^*,Z^0W^\pm W^{\mp *}, Z^0 Z^0Z^{0*}$, in
the intermediate mass range.
For the case $e^+e^-\ar e^+e^-
H^0 \ar e^+e^- b\bar b$, see ref.~\cite{Boos}. While
ref.~\cite{Orange3} contains a very complete analysis of various
`exclusive' signals and backgrounds over the whole allowed range
of $M_{H^0}$. This study is
based on the complete tree-level computation of the
processes $e^+ e^-\ar \ell^+\ell^- b\bar b$ (with $\ell=e$ or $\mu$)
and $e^+ e^-\ar \ell_1\ell_2V_1V_2$ (where $\ell_1,\ell_2$ represent
$e$- and $\mu$-leptons and neutrinos, whereas $V_1V_2$ indicate the massive
electroweak gauge bosons $W^\pm$ and $Z^0$).\par
Our results are presented in tab.~I and in figs.~2-4. In order to keep all
our matrix elements safe from singularities we have implemented the
following cuts: $M_{f\bar f^{(')}}> 10$ GeV for all flavours $f^{(')}$, plus
$|\cos\theta_{e,\nu_e}|<0.95$ and $E_{e,\nu_e}>10$ GeV for
electrons/positrons and corresponding neutrinos. These cuts
shouldn't affect the consistency of the analysis since, on the one hand,
the invariant mass $M_{f\bar f^{(')}}$ for the signals is always
$\approx M_{H^0},M_{W^\pm}$ or $M_{Z^0}$
and, on the other hand, the region along the beam pipe and the one
with soft energy are naturally restricted by the requirements of the
detectors.\par
Fig.~2 shows the distribution in missing-mass
$d\sigma/dM_{\rm{miss}}$, where $M_{\rm{miss}}^2=[(p_{e^+}+p_{e^-})-
p_{Z^0}]^2$, for  processes (\ref{proc1})--(\ref{proc3})
summed together, for the selection of Higgs masses
$M_{H^0}=110, 125, 140$ and 155 GeV,
at the CM energies $\sqrt s = 300$
and 500 GeV.
We have not included here any BR for the massive vector boson $Z^0$.
It is clear from these plots that the prospects of disentangling
the Higgs boson in the mass range 110 GeV
$\Ord$ \MH\ $\Ord$ 155 GeV remain quite promising even in the presence
of all the backgrounds coming from the non Higgs resonant diagrams of processes
(\ref{proc2})--(\ref{proc3}), summed over all $\ell$- and
$q$-flavour combinations,
if a mass resolution of $\approx15$ GeV
or better can be achieved (bins in fig.~2 are 5 GeV wide).\par
In  process (\ref{proc2}) the background is relevant in the $M_{\rm{miss}}$
spectrum only for $M_{f\bar f'
W^\pm}\OOrd 2 M_{W^\pm}\approx 160$ GeV, since here the main contribution
comes from the triple  vector boson production
$Z^0W^\pm W^{\mp *}$ (with the subsequent
decay $W^{\pm *}\ar f\bar f'$, diagrams \# 7--9, 18--20, 23--26 of
fig.~1b). Therefore, in \eeffwz\ both the
diagrams containing $Z^0Z^{0*}$ production diagrams (followed by $Z^{0*}\ar
f\bar f$ with a successive $W^\pm$-bremsstrahlung, graphs \# 12--15
with $Z^0$ propagator, or by $Z^{0*}\ar W^\pm W^{\mp *}\ar W^\pm f\bar
f'$, diagrams \# 23--24 in  fig.~1b) and the
non resonant diagrams (\# 1--6, 10--11, 12--15 with $\gamma$ propagator,
16--17
and 21--22 of  fig.~1b), as well all the interferences, are
negligible.
For process (\ref{proc3}) the shape is rather flat in the region \MH $\OOrd
110$ GeV, with the main contribution here coming from
$Z^0Z^0\gamma^*$ production (followed by $\gamma^*\ar
f\bar f$, graphs \# 4--6 with  $\gamma$ propagator, in fig.~1c),
whereas all the
other background contributions (i.e. the $Z^{0*}\ar f\bar f$ resonant diagrams
\# 4--6 with  $Z^0$ propagator, and the $H^0\ar f\bar f[Z^0Z^0]$
graphs \# 13--14[15--17], as well all the interferences) are
quite small.
Since for processes (\ref{proc2}) and (\ref{proc3}) the quantity
$M_{\rm{miss}}$ has a natural minimum at $M_{W^\pm}$ and $M_{Z^0}$,
respectively, we cannot properly consider here the case of the Higgs peak
overlapping with the $Z^0$ one (i.e. $M_{H^0}\approx90$ GeV).
To do this, we should consider five particle final states
(with the substitutions $W^\pm\lra f\bar f'$ and $Z^0\lra f\bar f$
in (\ref{proc2})--(\ref{proc3})),
which are beyond our intentions. From these two processes, however,
we expect completely negligible rates in the  $Z^0$-region
(see the rapidly falling shape of the distribution
at small values of $M_{\rm{miss}}$, especially at $\sqrt s=500$ GeV,
in fig.~2).
Process (\ref{proc1}), especially important in the
case $M_{H^0}\approx M_{Z^0}$,
has been carefully studied in ref.~\cite{GHS}, and
we do not repeat here the same discussion.
\par
The strength of the couplings of the $H^0$ to the $Z^0$ and to the
$W^\pm$ can be
deduced by the magnitude of the cross sections (in the bremsstrahlung
and fusion processes), whereas the ones to (some of) the fermions
are measurable (at least relative to, e.g. \bbb) through the BRs,
by singularly selecting the various Higgs decay channels \cite{BCDKZ}.
If one would like to verify the expected spin and parity of the \sm\
Higgs boson then he should turn to study, e.g. the spectrum of the
cosine of the angle
of the $Z^0$ with respect to the beam direction, i.e. $\cos\theta_Z$. We
know that in the case of production of a scalar boson in  association
with a vector boson, the distribution in this angle tends to approach
the $\sin\theta_Z^2$ law at high energies \cite{BCDKZ,angular}.
However, since the Bjorken
Higgs contribution is
quite small if compared to the total sample of
events  (\ref{proc1})--(\ref{proc3}) (see tab.~I),
this dependence is almost completely
washed out. In fact, fig.~3 shows that the main contribution is due to
the \eeffz\ process through $Z^0Z^0$-production, which has the largest
cross section. The shape is peaked in the very forward
region, reflecting the $t$-channel exchange of a fermion.
\par
However, suitable cuts in
$M_{\rm{miss}}$ around \MH\ (e.g. $|M_{\rm{miss}}-M_{H^0}|<15$ GeV)
and in $E_{Z^0}$ around $E_{\rm{ave}}$ (e.g.
$|E_{Z^0}-E_{\rm{ave}}|<12.5$ GeV)
get rid of the most part of the
backgrounds, practically keeping all the signals.
Fig.~4 shows the dependence of the inclusive cross section
on $\cos\theta_Z$ once the above cuts in $M_{\rm{miss}}$ and  $E_{Z^0}$
are implemented. The $\sin\theta_Z^2$ law stands
out now quite clearly, especially in the central region
of the spectrum and for $\sqrt s=500$ GeV. At  $\sqrt s=300$ GeV
the backgrounds are still quite effective.
 From fig.~4 it is however clear that a cut
in the angle
of the $Z^0$-direction with respect to the beam, say, $|\cos\theta_Z|<0.8$,
should be quite successful in improving the signal-to-noise
ratio, thus allowing for high precision measurements of the Higgs boson
parameters.

\subsection*{4. Conclusions}

In summary, we have computed at tree-level integrated and differential
rates for the reactions
\eeffz, \eeffwz\ and \eeffzz\ at NLC CM energies ($\sqrt s=300$, 500
GeV), for all possible combinations of flavours $f^{(')}=u,d,s,c,b,
e,\nu_e,\mu,\nu_\mu,\tau,\nu_\tau$.
These processes involve
the production of the \sm\ Higgs boson $H^0$ through the Bjorken
channel \eehz, followed by $H^0\ar b\bar b$, $H^0\ar W^\pm
W^{\mp *}\ar  W^\pm f\bar f'$ and   $H^0\ar Z^0Z^{0*}\ar Z^0f\bar f$,
respectively (the main Higgs decay channels in the
intermediate mass range), and the corresponding (both reducible and
irreducible) backgrounds. In particular, we focused our
attention on the interval
110 GeV $\Ord M_{H^0}\Ord $ 155 GeV.
It is extremely important for these mass values to exploit all the
possible Higgs search strategies and to know all the corresponding
backgrounds, since it is not clear
whether the intermediate mass range can be successfully covered by the
LHC, and at the same time it is beyond the possibilities of LEP II. In this
study we have assumed
an `inclusive' analysis in missing-mass, which
does not exploit any form of tagging
on the $H^0$-decay products but only on the ones of the $Z^0$ produced
in the Bjorken bremsstrahlung
reaction.
Although definite numerical
results can be given only after a proper experimental
simulation (including kinematical cuts, detector efficiencies,
hadronization, etc ...), we conclude that, via such analysis,
Higgs signals from the Bjorken channel
should be clearly
detectable for reasonable mass and energy resolutions, and
studies of the properties of the $H^0$
should be feasible, already for the standard integrated
luminosity of 10 fb$^{-1}$ per year, particularly at $\sqrt s=500$ GeV.

\subsection*{Acknowledgments}

We are grateful to A.G.~Akeroyd for useful comments and for reading
the manuscript.

\vfill
\newpage
\thispagestyle{empty}

\subsection*{Table Captions}

\begin{description}

\item[table~I] Cross sections for the signal ($e^+e^-\ar H^0Z^0\ar
XZ^0$) and for the complete processes
(\ref{proc1})--(\ref{proc3})  ($e^+e^-\ar XZ^0$), these latter summed together,
and with/without cut in $M_{\rm{miss}}$,
for $M_{H^0}=110,125,140,155$ GeV,
at $\sqrt s = 300$
and 500 GeV. The sum over all possible combinations of flavours in
eqs.~(\ref{proc1})--(\ref{proc3}) is implied.
The underlying cuts $M_{f\bar f^{(')}}\ge 10$ GeV,
$|\cos\theta_{e,\nu_e}|<0.95$ and $E_{e,\nu_e}>10$ GeV are implemented.
The BR of the reconstructed $Z^0$ is not included.

\end{description}

\subsection*{Figure Captions}

\begin{description}

\item[figure~1 ] Feynman diagrams contributing at lowest order to the processes
\eeffz\ (a),
$e^+e^-\rightarrow D\bar U W^+Z^0$ (b) and \eeffzz\ (c), where $(D,\bar
U)=(d,\bar u), (s,\bar c), (\mu,\bar\nu_\mu),
(\tau,\bar\nu_\tau)$ and $f=u,d,s,c,b,\mu,\tau,\nu_\mu,\nu_\tau$.
The diagrams for $e^+e^-\rightarrow U\bar D W^- Z^0$ are not shown,
but they can be trivially obtained from the $U\bar D W^- Z^0$ ones.
Internal wavy lines represent a $\gamma$, a $Z^0$ or a $W^\pm$,
as appropriate. Internal dashed lines represent the Higgs boson.

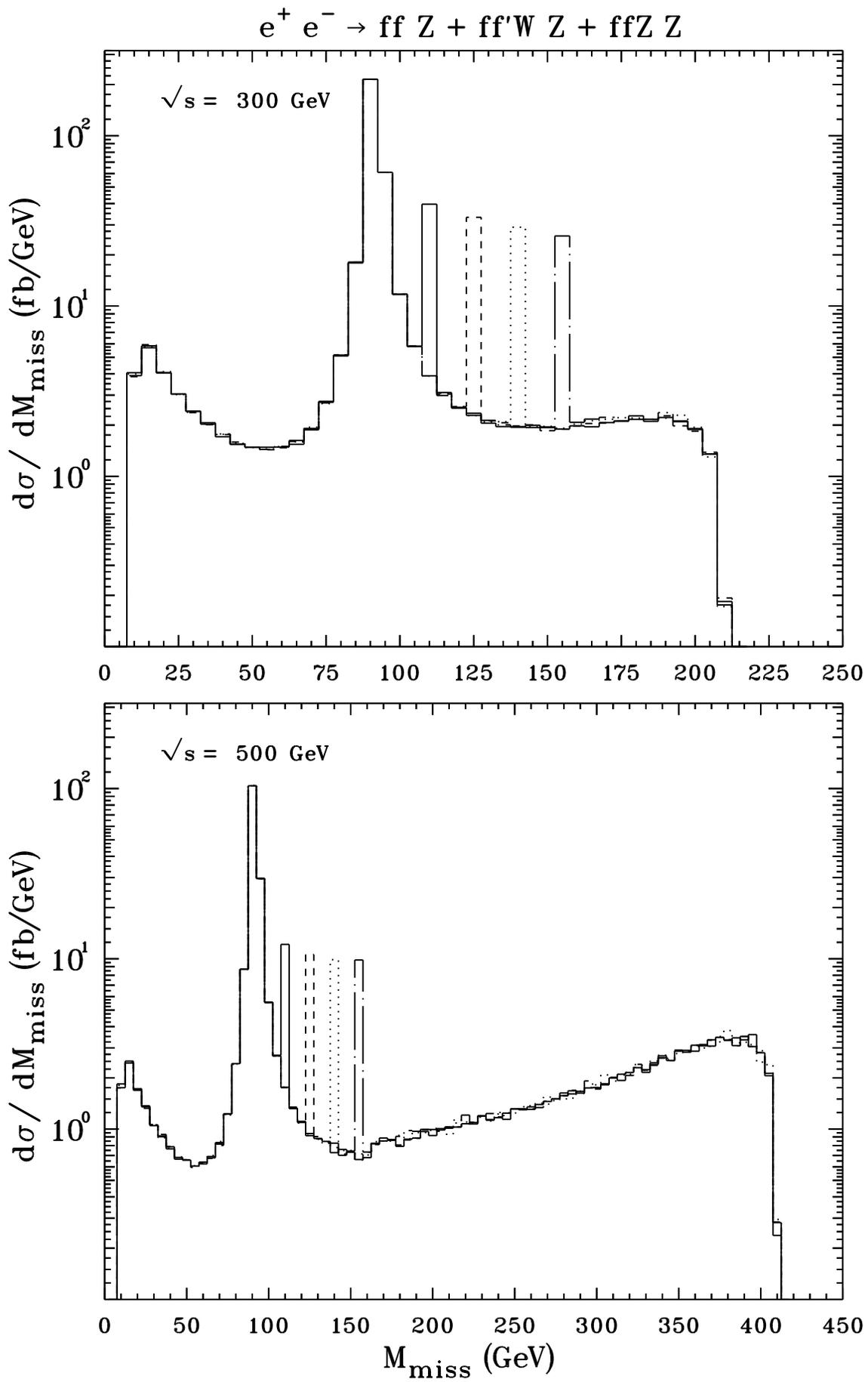
\item[figure~2 ] The differential distribution in missing-mass
$d\sigma/dM_{\rm{miss}}$, for  processes (\ref{proc1})--(\ref{proc3})
summed together (for all the possible combination
of flavours $f$ and $f'$), for
$M_{H^0}=110$ GeV (continuous line),
$M_{H^0}=125$ GeV (dashed line),
$M_{H^0}=140$ GeV (dotted line),
$M_{H^0}=155$ GeV (chain-dotted line),
at $\sqrt s = 300$
and 500 GeV. The underlying cuts $M_{f\bar f^{(')}}\ge 10$ GeV,
$|\cos\theta_{e,\nu_e}|<0.95$ and $E_{e,\nu_e}>10$ GeV are implemented.
The BR of the reconstructed $Z^0$ is not included.

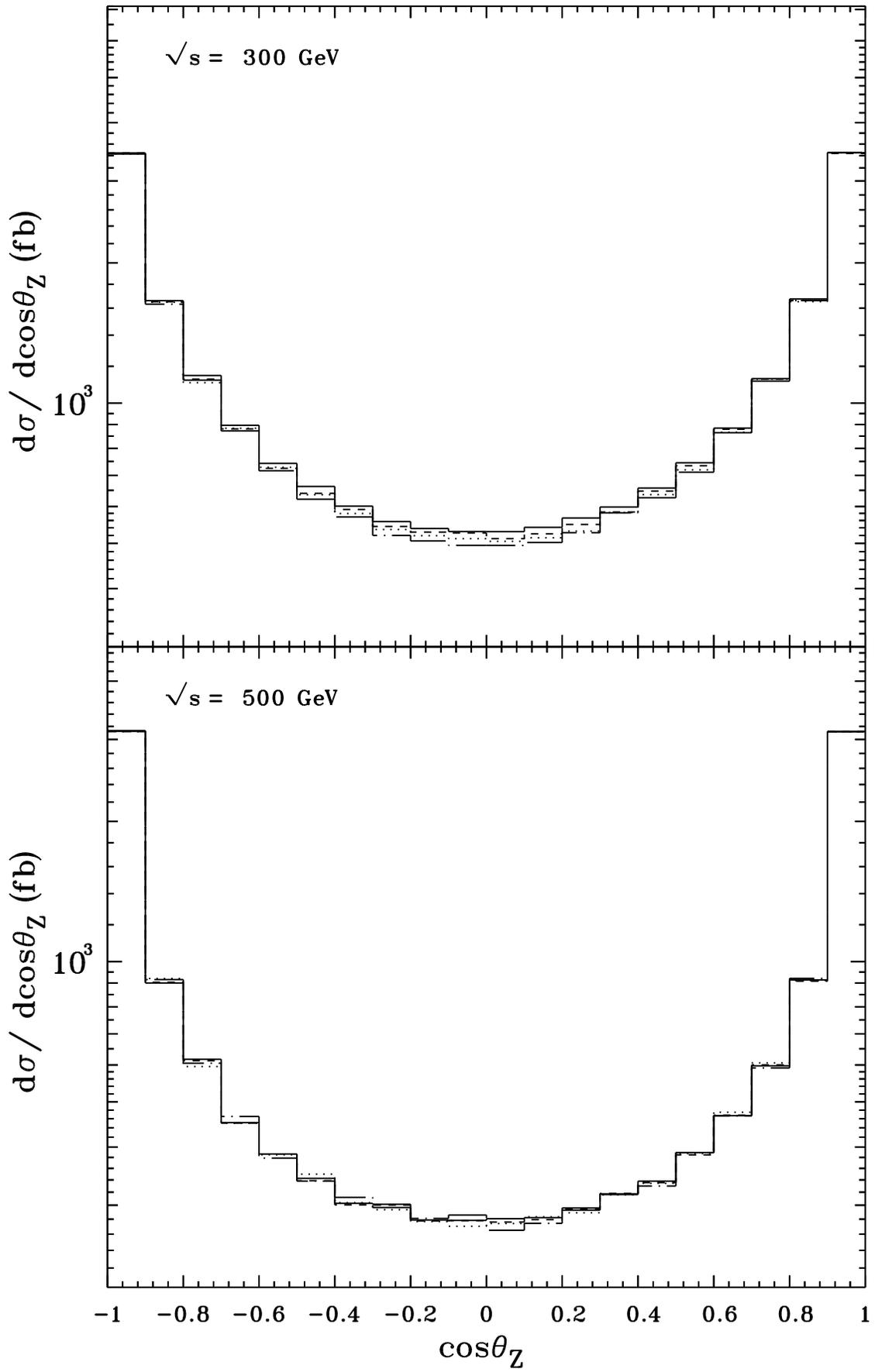
\item[figure~3 ] The differential distribution in the cosine of the
angle
of the $Z^0$ with respect to the beam direction $\cos\theta_Z$,
for processes (\ref{proc1})--(\ref{proc3})
summed together (for all the possible combination
of flavours $f$ and $f'$), for
$M_{H^0}=110$ GeV (continuous line),
$M_{H^0}=125$ GeV (dashed line),
$M_{H^0}=140$ GeV (dotted line),
$M_{H^0}=155$ GeV (chain-dotted line),
at $\sqrt s = 300$
and 500 GeV. The underlying cuts $M_{f\bar f^{(')}}\ge 10$ GeV,
$|\cos\theta_{e,\nu_e}|<0.95$ and  and $E_{e,\nu_e}>10$ GeV are implemented.
The BR of the reconstructed $Z^0$ is not included.

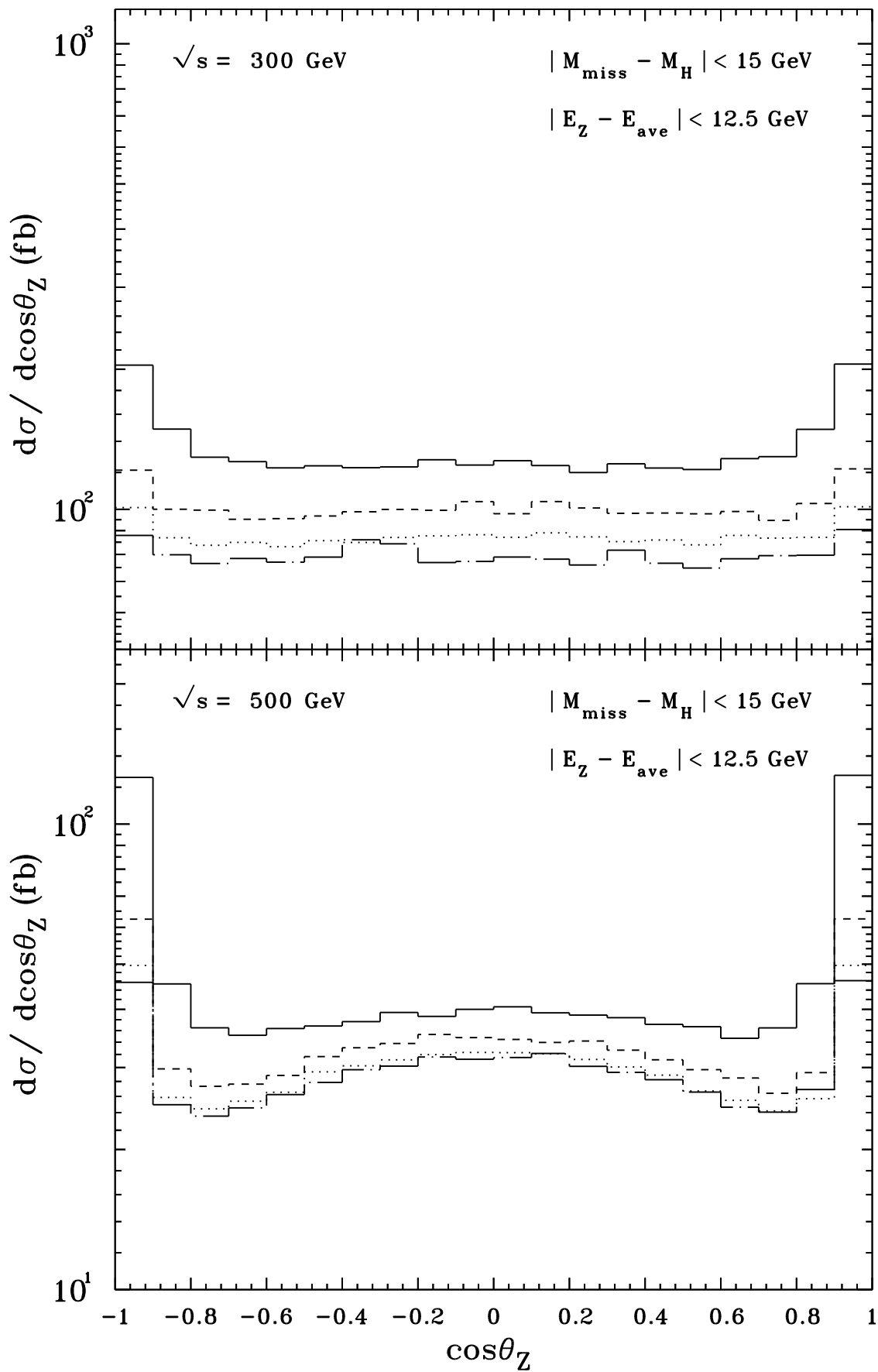
\item[figure~4 ] Same as fig.~3, after the cuts
$|M_{\rm{miss}}-M_{H^0}|<15$ GeV and $|E_{Z^0}-E_{\rm{ave}}|<12.5$ GeV.

\end{description}

\vfill
\newpage
\thispagestyle{empty}
\begin{table}
\begin{center}
\begin{tabular}{|c|c|c|c|}
\hline
\multicolumn{4}{|c|}
{\rule[-0.5cm]{0cm}{1.1cm}
$\sigma~{\rm{(fb)}}$}
 \\ \hline
\rule[-0.5cm]{0cm}{1.1cm}
$M_{H^0}~{\rm{(GeV)}}$
&  $e^+e^-\ar H^0Z^0\ar XZ^0$
&  $e^+e^-\ar XZ^0$
&  $e^+e^-\ar XZ^0$  \\ \hline\hline
\multicolumn{4}{|c|}
{\rule[-0.5cm]{0cm}{1.1cm}
 $~~\sqrt s=300(500)~{\rm{GeV}}$ }
\\ \hline
\rule[-0.6cm]{0cm}{1.3cm}
$110$ & $204(59)$ & $2154(1397)$ & $65(25)$    \\ 
\rule[-0.6cm]{0cm}{.9cm}
$125$ & $184(58)$ & $2131(1394)$ & $35(15)$     \\  
\rule[-0.6cm]{0cm}{.9cm}
$140$ & $163(56)$ & $2116(1394)$ & $31(12)$     \\  
\rule[-0.6cm]{0cm}{.9cm}
$155$ & $140(54)$ & $2101(1407)$ & $30(11)$     \\  \hline\hline
\rule[-0.6cm]{0cm}{1.1cm}
{}~~ & no $M_{\rm{miss}}$ cut & no $M_{\rm{miss}}$ cut &
$|M_{\rm{miss}}-M_{H^0}|<15$ GeV      \\  \hline\hline
\multicolumn{4}{|c|}
{\rule[-0.5cm]{0cm}{1.1cm}
$M_{f\bar f^{(')}}\ge 10$ GeV\quad\quad\quad\quad
$|\cos\theta_{e,\nu_e}|<0.95$\quad\quad\quad\quad
$E_{e,\nu_e}>10$ GeV}
 \\ \hline
\multicolumn{4}{c}
{\rule{0cm}{.9cm}
{\Large Table I}}  \\
\multicolumn{4}{c}
{\rule{0cm}{.9cm}}

\end{tabular}
\end{center}
\end{table}

\vfill

\end{document}